\documentclass[journal]{IEEEtran}

\usepackage{epsfig,amsmath,amstext,amssymb,mymacros}
\usepackage{graphicx,latexsym}
\usepackage{cite}
\usepackage{acronym}
\usepackage{color}
\usepackage{wrapfig}
\usepackage{epsfig}
\usepackage{epstopdf}
\usepackage{placeins}
\usepackage[caption=false]{subfig}
\usepackage{url}

\begin{document}
\pagenumbering{gobble}

\title{Anderson localization in a partially random Bragg grating and a conserved area theorem}

\author{Arash Mafi\\
{\small\em
Department of Physics \& Astronomy and Center for High Technology Materials, University of New Mexico, Albuquerque, NM 87131, USA
}}

\maketitle

\section*{Abstract}
We investigate the gradual emergence of the disorder-related phenomena in intermediate regimes
between a deterministic periodic Bragg grating and a fully random grating and highlight two critical 
properties of partially disordered Bragg gratings. First, the integral of the logarithm of the transmittance over the reciprocal 
wavevector space is a conserved quantity. Therefore, adding disorder merely redistributes the transmittance over the 
reciprocal space. Second, for any amount of disorder, the average transmittance decays exponentially with the 
number of grating layers in the simple form of $\exp(-\eta N)$ for sufficiently large $N$, where $\eta$ is a constant 
and $N$ is the number of layers. Conversely, the simple exponential decay form does not hold for small $N$ except for 
a highly disordered system. Implications of these findings are demonstrated.

\section{Introduction}
An optical Bragg grating is created by periodic variations in the permittivity of dielectric material. 
Depending on the size of the periodicity, the grating can act as a highly reflective (bandgap) or a highly 
transmittive (bandpass) mirror~\cite{SalehTeich}. A bandgap or bandpass is not limited to an optical Bragg grating: 
in solid-state physics, a periodic lattice of atoms or molecules in a crystal modulates the background potential for electron
wave function and similar badgap and bandpass characteristics are observed. Understanding the electronic band 
structure is at the foundation of solid-state physics.

In practice, no potential is perfectly periodic and no refractive index modulation is disorder-free. Therefore,
it is important to investigate the impact of disorder on the band structure in electronic systems, 
optical gratings, and other coherent wave systems~\cite{Ziman}. There have been numerous studies of disordered coherent wave 
systems, much of which follows the seminal work of Philip Anderson~\cite{Anderson}. He showed that for a sufficiently high level of 
disorder the electronic wavefunction is exponentially attenuated, resulting in diminished transport and 
conductivity~\cite{Anderson2}, as also confirmed in subsequent work by 
others~\cite{Erdos,John1,Lagendijk1,Chabanov,Billy,Lahini,Abouraddy,Schwartz,SalmanOL,SalmanNature,SegevNaturePhotonicsReview,MafiAOP}.

A one-dimensional (1D) periodic Bragg grating is simple and elegant--its band structure is easy to understand and 
analytical calculations can often be performed to study its characteristics~\cite{Pendry}. Moreover, many of its 
properties are generic and can be extended to 2D and 3D gratings. A highly disordered 1D grating made from a random 
stack of dielectric slabs has also been treated analytically by Berry and Klein~\cite{Berry}.
They showed that the average transmitted intensity drops exponentially with the number of slabs.  

The problem of a perfectly periodic Bragg grating is well understood, and so is that of a highly disordered grating. 
The intention of this article is to shed further light on intermediate situations with moderate disorder~\cite{Berry,Kondilis,Bliokh}. 
We explore the transition from a deterministic periodic Bragg grating to a fully random grating, and highlight 
the gradual emergence of the disorder-related phenomena in intermediate regimes.
In particular, we present two critical properties of partially disordered 1D Bragg gratings. Because our analysis 
is carried out for a stack of dielectric slabs similar to Ref.~\cite{Berry}, we present these properties in the 
language of dielectric slabs for simplicity:

{\em First} we find a globally conserved quantity in the reciprocal wavevector space that is quite instrumental in 
visualizing the wave localization behavior in a partially disordered stack of dielectric slabs. We show that
the total area under the curve of the logarithm of the transmittance plotted in the reciprocal wavevector space
is the same for all Bragg gratings, regardless of the amount of disorder. As a consequence of this conservation law,
if the transmittance is higher in some region of the reciprocal space, it has to be lower in another
region to conserve the total area. For example, the presence of a strong bandgap for a periodic Bragg grating necessitates 
a strong bandpass in a different region of the reciprocal space.
 
{\em Second} we observe that for any amount of disorder, the average transmitted intensity decays exponentially with 
the number of slabs as $\exp(-\eta N)$ for sufficiently large $N$. $\eta$ is a constant whose value depends 
on the amount of disorder, and $N$ is the number of slabs. Even in the absence of disorder, the simple exponential 
decay law holds in the bandgap region for large $N$. Conversely, the simple exponential decay form {\em does not} hold for 
small $N$ except for a highly disordered system.

In the following, these ideas will be presented in the framework of optical transmission through a 1D stack of dielectric 
slabs. The slabs are assumed to be identical with refractive index $n_2$ and are embedded in a background dielectric of 
refractive index $n_1$. The disorder is introduced by randomizing the location of the slabs.
\section{Background}
The transmission (transfer) matrix for a lossless reciprocal mirror (dielectric slab) can be expressed as
\begin{align}
M_t~=~ 
\begin{bmatrix}
      1/t^\ast        & r/t  \\[0.3em] \relax
      r^\ast/t^\ast   & 1/t \relax
\end{bmatrix}
~=~
\dfrac{1}{\tau}
\begin{bmatrix}
       1      & i \rho     \\[0.3em]
      -i\rho  & 1 
\end{bmatrix}.
\label{eq:generalM}
\end{align}
$t$ and $r$ are the amplitude transmittance and reflectance, respectively, 
and are functions of the dielectric constants and geometric properties of 
the slab and frequency and incidence angle of light~\cite{SalehTeich}. 
They also satisfy the losslessness condition $|t|^2+|r|^2=1$. The latter 
simplified form is obtained by formally incorporating a thin layer of the 
background $n_1$ dielectric in the $n_2$ slab, where $\tau=|t|$ and $\rho=|r|$.
The latter form is used in the following discussions as it does not affect 
the generalities of the presented arguments.

The transfer matrix of an array of $N$ {\em identical} dielectric slabs with  
varying separations (gaps) of the background dielectric material can be expressed as
\begin{equation}
{{\rm M}^{(N)}}=\prod_{n=1}^N \big(M_t.M^{(n)}_\varphi\big),\
\quad\
M^{(n)}_\varphi=
\begin{bmatrix}
       e^{i\varphi_n} & 0               \\[0.3em]
       0              & e^{-i\varphi_n} 
\end{bmatrix},
\end{equation}
where $M^{(n)}_\varphi$ is the translation matrix accounting for the phase accumulation $\varphi_n$ 
in the $n$th gap.
Here, $M^{(N)}_\varphi$ is assumed to be the identity matrix. The total wave transfer matrix of 
the system is given by
\begin{equation}
{{\rm M}^{(N)}}=
\dfrac{1}{\tau^N}
\prod_{n=1}^N
\begin{bmatrix}
           e^{i\varphi_n}    &  i \rho e^{-i\varphi_n}    \\[0.3em]
   -i \rho e^{i\varphi_n}    &         e^{-i\varphi_n} 
\end{bmatrix},
\label{eq:totalBmatrix}
\end{equation}
where the total intensity transmittance is given by the $(2,2)$ element of the ${{\rm M}^{(N)}}$: 
${\cal T}(N,\tau;\{\varphi_n\})=|{{\rm M}_{22}^{(N)}}|^{-2}$.  
\section{Transmission through a periodic vs. random stack}
\begin{figure}[t]
\centerline{\includegraphics[width=.9\columnwidth]{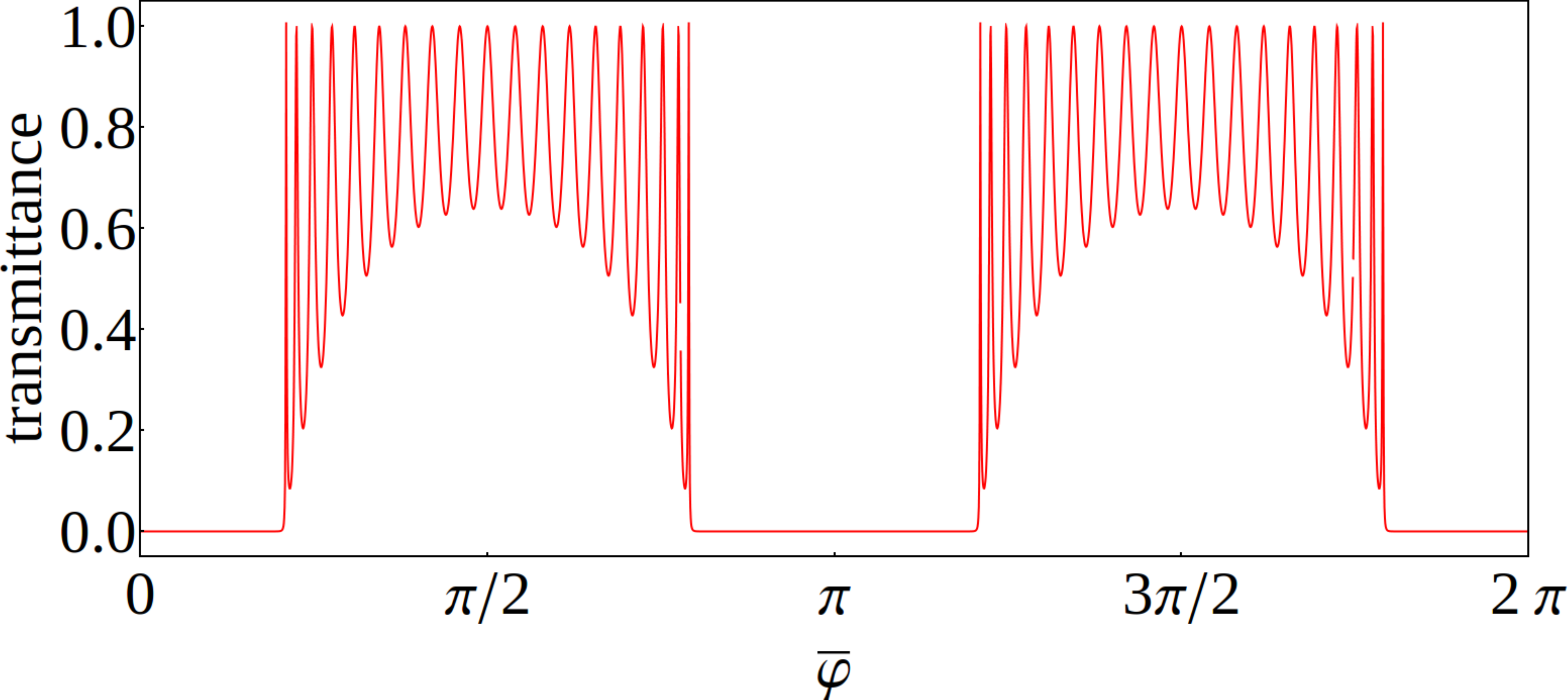}}
\caption{\label{fig:Bragg-tau-80-N-20}Transmittance is plotted as a function of the gap phase for a 20-layer periodic
Bragg grating with $\tau=0.8$.}
\end{figure}

The most common application of the preceding analysis is to study periodic Bragg gratings.
For a periodic grating where the gap thickness between consecutive slabs is identical, 
the accumulated phase in each gap takes a common value $\bar{\varphi}$. 
In Fig.~\ref{fig:Bragg-tau-80-N-20}, the total transmittance ${\cal T}$ is plotted 
for a {\em periodic} stack of $N=20$ dielectric mirrors with $\tau=0.8$ as a 
function $\bar{\varphi}$, where one can clearly see the familiar bandpass and 
bandgap regions.

In practice, it is impossible to maintain a uniform gap (fixed $\bar{\varphi}$)
between all slabs, and some randomness is inevitable. A convenient way to parametrize 
a partially disordered grating is to assume that the accumulated phase in each gap
is a random number with an average value of $\bar{\varphi}$ with some probability
for variation around the average. We adopt the definition in Eq.~\ref{eq:partialrandom}, 
which states that $\varphi_n$ is chosen from a uniform random distribution between 
$\bar{\varphi}-\alpha\pi$ and $\bar{\varphi}+\alpha\pi$. The disorder level is 
parametrized by $\alpha$: the {\em deterministic} periodic 
grating is identified by $\alpha=0$ and a fully random grating by $\alpha=1$.
\begin{equation}
\varphi_n=\bar{\varphi}+\theta_n, \qquad
\theta_n\in \alpha\times{\rm unif}[-\pi,\pi],\ \ \ 0\le\alpha\le 1.
\label{eq:partialrandom}
\end{equation}

\begin{figure}[t]
\centerline{\includegraphics[width=.85\columnwidth]{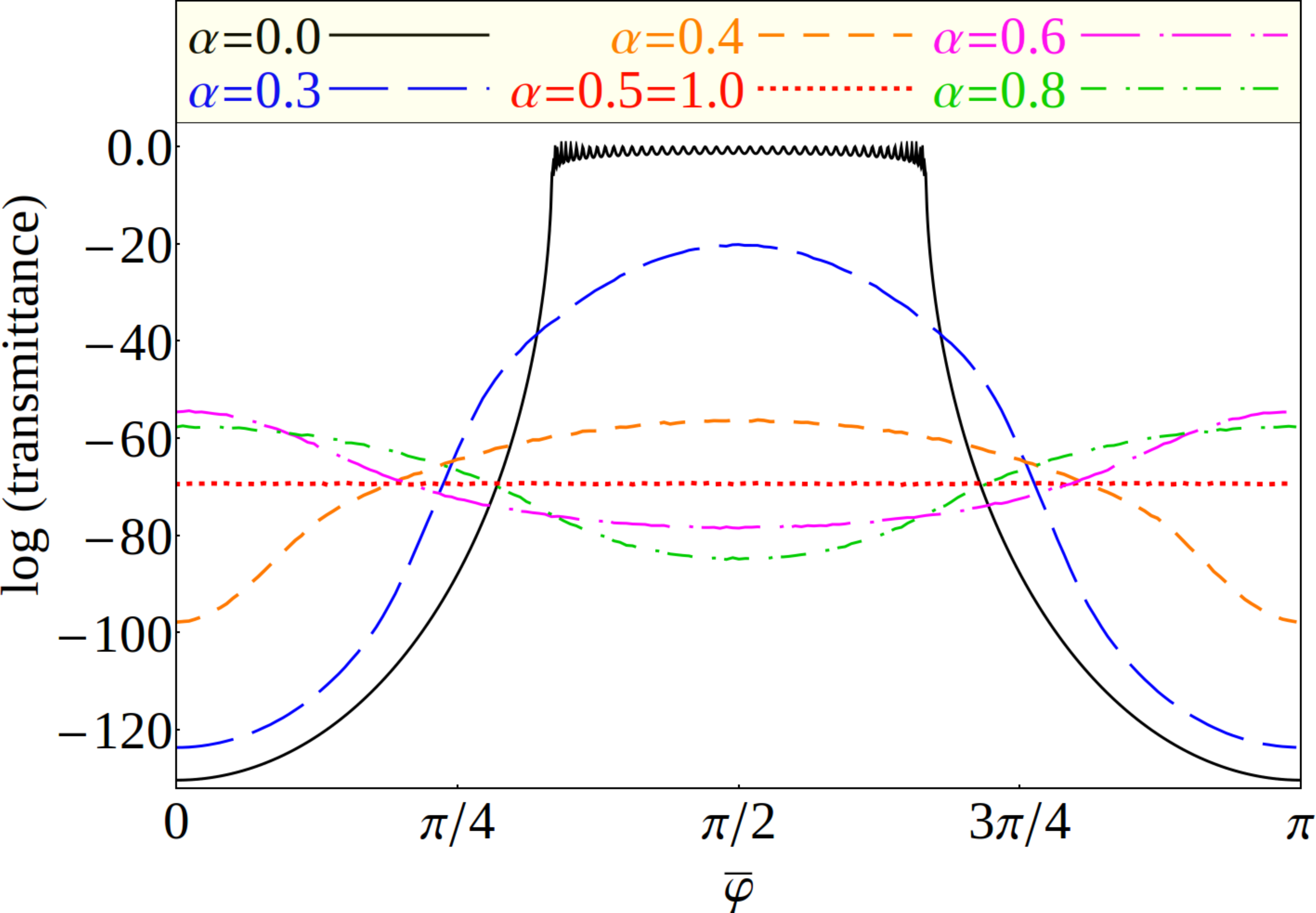}}
\caption{\label{fig:bragg-partial-random}Average transmittance is plotted in logarithmic scale as a function of the 
average gap phase. Each grating has $N=50$ layers with $\tau=0.5$. Different curves 
are for different values of the disorder parameter $\alpha$. The area under each curves is the same.}
\end{figure}

The transmittance for different levels of randomness characterized by $\alpha$ is 
plotted in Fig.~\ref{fig:bragg-partial-random} as a function of $\bar{\varphi}$.
Here, $N=50$ and $\tau=0.5$ is assumed and the vertical scale is logarithmic. 
For each value of $\alpha$, the transmittance curve is {\em properly averaged}
(proper averaging will be explained shortly), for an ensemble of 10,000 different 
gratings, each for a different set of $\{\varphi_n\}$ values. 

For a periodic Bragg grating with $\alpha=0$, an analytical formula exists for transmittance:
\begin{align}
\label{eq:BraggAnalytical}
&{\cal T}(N,\tau,\bar{\varphi})=\dfrac{\tau^2}{\tau^2+\rho^2\Psi^2_N},\\
\nonumber
&\Psi_N=\dfrac{\sin(N\Phi)}{\sin(\Phi)},\qquad \Phi=\cos^{-1}\Big(\dfrac{\cos(\bar{\varphi})}{\tau}\Big). 
\end{align}
The transmittance curve in Fig.~\ref{fig:Bragg-tau-80-N-20} and the black solid curve in Fig.~\ref{fig:bragg-partial-random} 
follow Eq.~\ref{eq:BraggAnalytical}. The oscillatory bandpass regions correspond to
$|\cos(\bar{\varphi})|<\tau$ where $\Phi$ is real, while the bandgap region correspond to $|\cos(\bar{\varphi})|>\tau$,
where $\Phi$ is imaginary. In the bandgap region, it is more convenient to redefine $\Psi_N$ and $\Phi$ as
\begin{equation}
\label{eq:BraggAnalytical2}
\Psi_N=\dfrac{\sinh(N\Phi)}{\sinh(\Phi)},\qquad \Phi=\cosh^{-1}\Big(\dfrac{\cos(\bar{\varphi})}{\tau}\Big). 
\end{equation}

For a partially random grating with $\alpha>0$, the transmittance curve would look rather noisy. The smooth transmittance curves 
in Fig.~\ref{fig:bragg-partial-random} are only obtained after properly averaging over many gratings.
Because the transmittance is a multiplicative quantity, {\em proper averaging} of the random transmitted intensity is 
done by averaging the logarithm of the transmittance~\cite{Anderson2,Berry}. Using Eq.~\ref{eq:totalBmatrix}, we can formally express 
the averaging as
\begin{align}
\nonumber
\log\Big({\cal T_{\rm avg}}\Big)
&=
\Big(
\prod_{n=1}^N
\int^{\bar{\varphi}+\alpha\pi}_{\bar{\varphi}-\alpha\pi}
\dfrac{d\varphi_n}{2\pi\alpha}\Big)
\log\Big({\cal T}(\{\varphi_n\})\Big)\\
&=2N\log(\tau)
+\Omega(N,\tau,\bar{\varphi},\alpha),
\end{align}
where $2N\log(\tau)$ comes from the overall multiplicative factor in Eq.~\ref{eq:totalBmatrix} and 
$\Omega(\tau,N,\{\varphi_n\})$ comes from averaging the matrix multiplication part. 

For the case of a {\em totally random} dielectric stack with $\alpha=1$, Berry 
and Klein~\cite{Berry} have rigorously shown that $\Omega(N,\tau,\bar{\varphi},1)=0$;
therefore, ${\cal T}_{\rm avg}=\exp(2N\log(\tau))$ and is independent 
of the value of $\bar{\varphi}$. The numerical simulation of Fig.~\ref{fig:bragg-partial-random} 
in red dotted line agrees with the analytical results, where 
$\log({\cal T}_{\rm avg})=2N\log(\tau)=-69.31$ for $N=50$ and $\tau=0.5$.

For intermediate values of $\alpha$ between $0$ and $1$, a closed-form solution is not available 
for ${\cal T}_{\rm avg}$ and one must resort to numerical plots similar to those presented in Fig.~\ref{fig:bragg-partial-random}. 
However, all these curves follow a remarkable conservation law, which is a consequence of 
$\int^\pi_0d\bar{\varphi}~\Omega=0$.
It can be verified that by changing the amount of randomness via $\alpha$, 
the averaged transmittance profile redistributes itself over the mean accumulated gap phase $\bar{\varphi}$
in a such way that it preserves the area under the logarithm-transmittance curves in Fig.~\ref{fig:bragg-partial-random}:
\begin{equation}
{\cal A}=\int^{\pi}_0 d\bar{\varphi}\ \log\Big({\cal T_{\rm avg}}(N,\tau,\bar{\varphi},\alpha)\Big)=2\pi N\log(\tau).
\end{equation}

The area is independent of the randomness level $\alpha$ and is the same for all curves. 
This {\em conserved transmittance area theorem} gives special status to the case of a
totally random dielectric stack marked with $\alpha=1$ and the conventional periodic Bragg 
grating marked with $\alpha=0$. The totally disordered grating incorporates all possible 
values of phase and $\bar{\varphi}$ loses its special standing; therefore, the 
transmittance becomes independent of $\bar{\varphi}$ by uniformly spreading the available 
conserved area ${\cal A}$ over all values of $\bar{\varphi}$. Conversely, the periodic Bragg grating
with $\alpha=0$ provides the most nonuniform distribution of the available conserved area 
${\cal A}$ over the space of $\bar{\varphi}$ with highly depressed values of transmittance 
in the bandgap, accompanied by a large transmittance in the bandpass to compensate.

In Appendix, we offer a proof of the conserved area theorem. It is shown that the conserved area
theorem applies to the transmittance for a general partially random grating and averaging
over $\{\theta_n\}$ is not required ($\{\theta_n\}$ was defined in Eq.~\ref{eq:partialrandom}). 
In the absence of ensemble averaging, the transmittance curve would look rather noisy but 
still satisfies the conserved area theorem. 

The conserved area theorem provides a very intuitive and useful approach to visualize the impact of partial disorder.
It is important to note that the accumulated phase values in the gaps are proportional to 
the wavevector $k$, where $\varphi_n=n_1kd_n$ and $d_n$ is the {\em random} thickness of the $n$th gap.
Each curve in Fig.~\ref{fig:bragg-partial-random} should be regarded as corresponding to a grating
with an {\rm average} gap thickness $\bar{d}$, where $\bar{\varphi}=n_1k\bar{d}$. Therefore,
the horizontal axis $\bar{\varphi}$ in Fig.~\ref{fig:bragg-partial-random} actually represents the wavevector,
and the conserved area theorem is a statement about the integral of the average of the logarithm of the 
transmittance over the reciprocal space.
\section{Transmittance scaling with the number of mirrors}
The exponential decay of transmittance (relative optical intensity) is regarded as the main 
signature of Anderson localization. For the case of a totally random stack marked with $\alpha=1$, 
it was previously shown that the average transmittance scales like ${\cal T_{\rm avg}}=\exp(-\eta N)$, 
where the exponent is proportional to the number of mirrors. The scaled exponent $\eta$ is independent 
of $N$ and is given by $\eta=-2\log(\tau)\ge 0$.

In this section, it is argued that neither the exponential decay nor its universal scaling with $N$ is
specific to the case of a totally random stack marked with $\alpha=1$. Rather, for a sufficiently 
large number of mirrors $N$, ${\cal T_{\rm avg}}$ always scales like $\exp(-\eta N)$ for all gratings. 
The only exception is for the periodic Bragg grating marked with $\alpha=0$ in the bandpass region
where $|\cos(\bar{\varphi})|<\tau$. 

\begin{figure}[t]
\centerline{\includegraphics[width=.85\columnwidth]{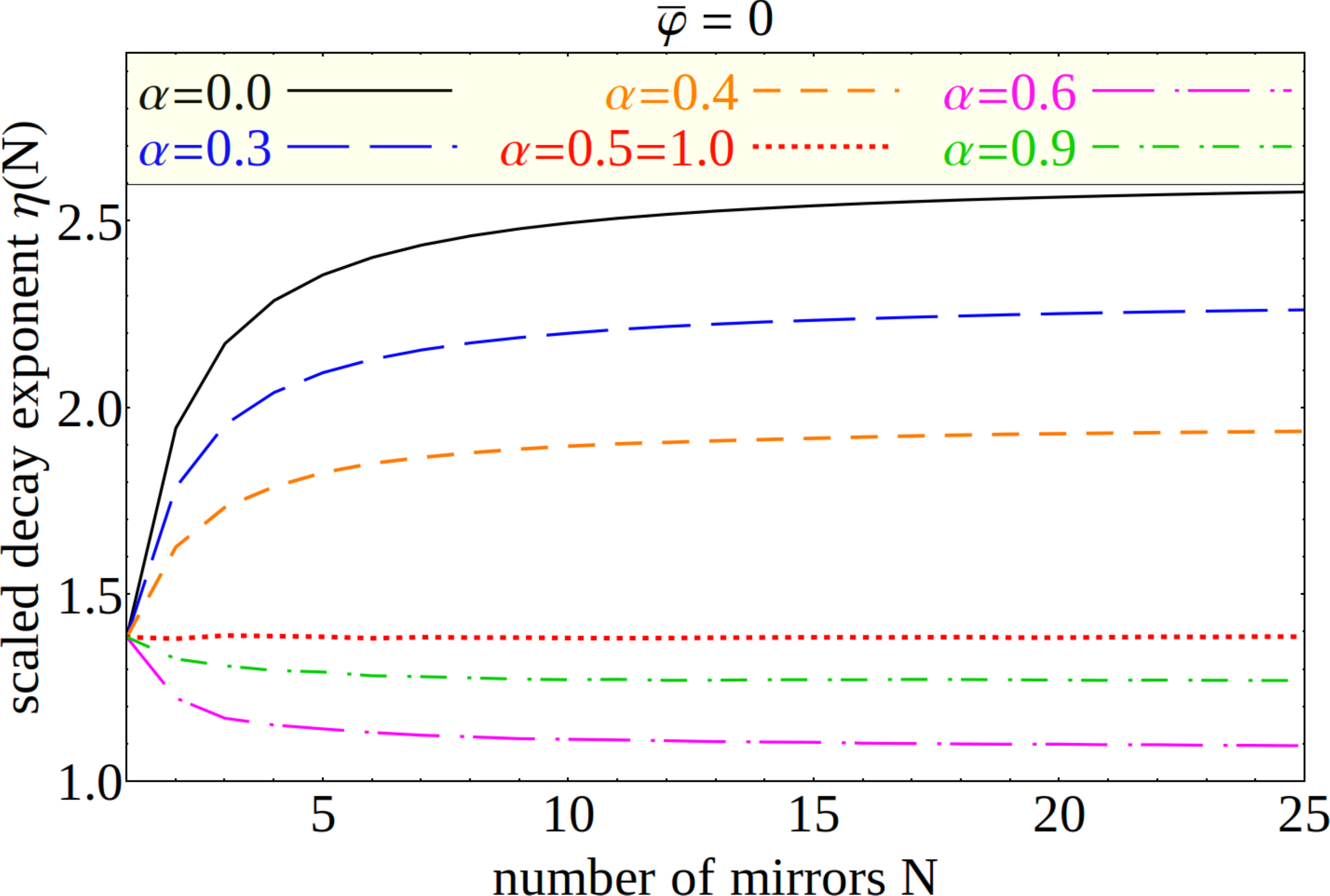}}
\caption{\label{fig:decay-plot-1}Average scaled decay exponent $\eta$ is plotted as a function of the number of mirrors $N$ for different
values of the disorder parameter $\alpha$. $\tau=0.5$ and $\bar{\varphi}=0$ are assumed here.
$\eta$ is meaningful only at integer values of $N$, but the points are connected for better visual clarity.}
\end{figure}

We start by studying the periodic Bragg grating with $\alpha=0$. According to Eq.~\ref{eq:BraggAnalytical}, the 
transmittance in the bandpass is a non-decaying oscillatory function of $N$. However, in the bandgap, it is an 
exponentially decaying function of $N$ for large $N$. This can been seen by using Eq.~\ref{eq:BraggAnalytical2} and noting that  
$\sinh(N\Phi)\sim\exp(N\Phi)/2$ for large $N$. After a few simple algebraic steps, it can be shown that
$\log({\cal T})\sim 2N\Phi$; therefore, $\eta = -2\Phi$ for large $N$. For $N=1$, Eq.~\ref{eq:BraggAnalytical}
shows that $\log({\cal T})= 2\log(\tau)$. In Fig.~\ref{fig:decay-plot-1}, $\eta$ is plotted as a function of 
$N$ for different values of $\alpha$, all for $\bar{\varphi}=0$ and $\tau=0.5$. The black solid line corresponds to $\alpha=0$, where 
$\eta$ starts at $2\log(\tau)\approx -1.37$ for $N=1$ and saturates at $\eta = 2\cosh^{-1}(1/\tau)\approx 2.63$ for large $N$. 

The plots of $\eta$ as a function of $N$ for different values of $\alpha$ follow a similar behavior to that of the 
periodic grating explained above. They all start at $\eta=2\log(\tau)$ for $N=1$ (see Eq.~\ref{eq:totalBmatrix}) and
asymptotically approach a limit for large $N$. Therefore, the average transmittance decays exponentially
for all cases at large $N$. In Fig.~\ref{fig:decay-plot-1}, which is specific to $\bar{\varphi}=0$,
the largest asymptotic value for $\eta$ is obtained for $\alpha=0$. This is not surprising, because
$\bar{\varphi}=0$ corresponds to the bottom of the bandgap of the periodic Bragg grating in 
Fig.~\ref{fig:bragg-partial-random}. 

\begin{figure}[t]
\centerline{\includegraphics[width=.85\columnwidth]{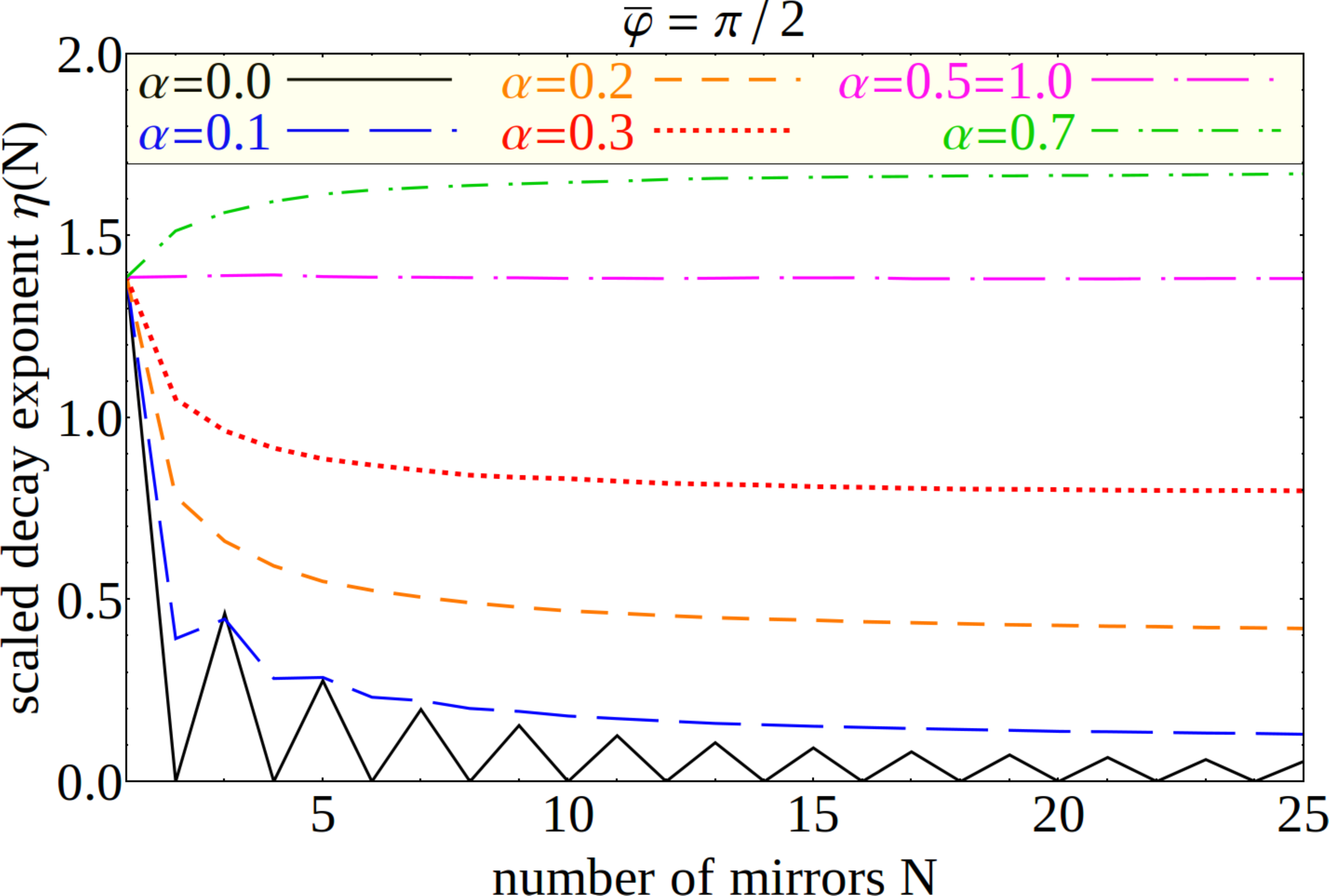}}
\caption{\label{fig:decay-plot-2}Similar to Fig.~\ref{fig:decay-plot-1}, except $\bar{\varphi}=\pi/2$.}
\end{figure}
A similar behavior is observed in Fig.~\ref{fig:decay-plot-2}, which is the same as Fig.~\ref{fig:decay-plot-1}, 
except $\bar{\varphi}=\pi/2$. The main difference is for the solid black line, which 
relates to the periodic Bragg grating with $\alpha=0$: $\eta$ oscillates and asymptotically approaches zero as $1/N$; therefore,
transmittance does not decay exponentially even for large $N$. This behavior is expected because $\bar{\varphi}=\pi/2$ is at the center of
the bandpass for the Bragg grating. When a slight amount of disorder marked by $\alpha=0.1$ is introduced, the $\eta$-curve
(long-dashed blue) goes through a brief oscillation and then approaches asymptotically to a nonzero value, indicating 
that for large $N$ the ${\cal T_{\rm avg}}=\exp(-\eta N)$ behavior is observed. Increasing randomness to $\alpha=0.5$
increases the scaled decay exponent monotonically. Increasing $\alpha$ beyond 0.5 initially increases $\eta$, but eventually
rolls it back to make the $\eta$-curve for $\alpha=1.0$  the same as that of $\alpha=0.5$. 

The universal exponential decay behavior at large $N$ is expected because all 1D disordered systems are Anderson localized~\cite{Anderson2}.
However, the converse observation that for small $N$, the transmittance curve can significantly deviate from the simple exponential 
decay form despite being Anderson-localized is worthy of special attention. The exponential decay is regarded as 
the main signature for Anderson localization and the stated converse observation signifies the importance of asymptotic analysis
in establishing Anderson localization. In other words, large deviations from an exponential decay in the first few layers
do not exclude the possibility of Anderson localization.
\section{Conclusions}
The transition from a deterministic periodic Bragg grating to a fully random grating and
the gradual emergence of the disorder-related phenomena in the intermediate regimes is studied 
in detail. It is shown that for any amount of disorder the average transmittance decays exponentially with the 
number of grating layers in the simple form of $\exp(-\eta N)$ for sufficiently large $N$, where $\eta$ is a 
constant. For a highly disordered grating, this simple form is true regardless of the number of mirrors $N$.
Conversely, for small $N$, the transmittance curve can deviate from the simple exponential decay form and 
can even oscillate as it decays. Whether increasing the disorder increases or decreases the decay exponent
depends on the {\em location} of the grating in the reciprocal wavevector space, the amount of disorder, and
other parameters defining the grating.

It is also shown that the integral of the logarithm of the transmittance over the reciprocal 
wavevector space is a conserved quantity. The randomness in the examples used in this article is
introduced through a uniform distribution for convenience. However, the conserved area theorem 
works for other methods of randomization such as Gaussian and Poisson distributions and is valid even
in the absence of ensemble averaging.
It is plausible that the conserved area theorem can be extended to 2D and 3D photonic crystals~\cite{Mafi}, which will be 
the subject of future studies.
\section{Appendix}
Here, we present a proof of the conserved area theorem, i.e. 
$\int^\pi_0 d\bar{\varphi} \log\Big(|{{\rm M}_{22}^{(N)}}|^{-2}\Big)=2N\log(\tau)$,
where ${{\rm M}^{(N)}}$ is defined in Eq.~\ref{eq:totalBmatrix}. Equivalently, we can show
$\int^\pi_0 d\bar{\varphi} \log\Big(|{{\rm P}_{22}^{(N)}}|^{-2}\Big)=0$, where
${{\rm P}^{(N)}}$ is defined as:
\begin{equation}
{{\rm P}^{(N)}}=\tau^N\times{{\rm M}^{(N)}}=
\left(
\prod_{n=1}^N
e^{i\varphi_n}
\right)
\times
\prod_{n=1}^N
\begin{bmatrix}
           1    &  i \rho e^{-2i\varphi_n}    \\[0.3em]
   -i \rho      &         e^{-2i\varphi_n} 
\end{bmatrix}.
\label{eq:totalBmatrix2}
\end{equation}
Using Eq.~\ref{eq:totalBmatrix2} and defining $z=\exp(-2i\bar{\varphi})$, it can be shown that
\begin{equation}
\int^\pi_0 d\bar{\varphi} \log\Big({{\rm P}_{22}^{(N)}}\Big)=
\gamma + \oint_C \dfrac{dz}{-2iz} \log\Big(1+\chi(z,\rho,\{\theta_n\})\Big),
\end{equation}
where $\gamma=i\int^\pi_0 d\bar{\varphi} \sum^N_{n=1}\phi_n$. $\chi$ is a holomorphic function of 
$z$ that vanishes at $z=0$; therefore, Cauchy's residue theorem implies that the contour 
integral vanishes and  
$\int^\pi_0 d\bar{\varphi} \log\Big({{\rm P}_{22}^{(N)}}\Big)=\gamma$. Therefore,
we can show that
\begin{equation}
\int^\pi_0 d\bar{\varphi} \log\Big(|{{\rm P}_{22}^{(N)}}|^{-2}\Big)=-\int^\pi_0 d\bar{\varphi} \log\Big({{\rm P}_{22}^{(N)}}\Big)
+c.c.=0,
\end{equation}
which completes the proof (c.c. stands for the complex conjugate). In the last step, we have used the fact that $\gamma$ is a 
purely imaginary number. Note that averaging over $\{\theta_n\}$ is not needed to prove the theorem; therefore, it is valid
for each random transmission curve.




\begin{thebibliography}{99}
\bibitem{SalehTeich}
B. E. A. Saleh and M. C. Teich, {\em Fundamentals of Photonics}, (Wiley, 2007).
\bibitem{Ziman}
J. M. Ziman, {\em Models of Disorder,} (Cambridge University, 1979).
\bibitem{Anderson}
P. W. Anderson, Phys. Rev. \textbf{109}, 1492 (1958).
\bibitem{Anderson2}
P. W. Anderson, D. J. Thouless, E. Abrahams, and D. S. Fisher, Phys. Rev. B \textbf{22}, 3519 (1980).
\bibitem{Erdos}
P. Erd\"{o}s and R.C. Herndon, Adv. Phys. \textbf{31}, 65 (1982).
\bibitem{John1}
S. John, Phys.\ Rev.\ Lett.\ \textbf{58}, 2486 (1987).
\bibitem{Lagendijk1}
A. D. Lagendijk, B. van Tiggelen, B., D. S. Wiersma, {Phys. Today} \textbf{62}, 24 (2009).
\bibitem{Chabanov}
A. A. Chabanov, A. Stoytchev, and A. Z. Genack, Nature \textbf{404}, 850 (2000).
\bibitem{Billy}
J. Billy, V. Josse, Z. Zuo, A. Bernard, B. Hambrecht, P. Lugan, D. Cl\'{e}ment, L. Sanchez-Palencia, P. Bouyer, and A. Aspect, 
Nature \textbf{453}, 891 (2008).
\bibitem{Lahini}
Y.~Lahini, A.~Avidan, F.~Pozzi, M.~Sorel, R.~Morandotti, D.~N.~Christodoulides, and Y.~Silberberg,
 Phys.\ Rev.\ Lett.\  \textbf{100}, 013906 (2008).
\bibitem{Abouraddy}
A. F. Abouraddy, G. Di Giuseppe, D. N. Christodoulides, and B. E. A. Saleh, Phys.\ Rev.\ A\ \textbf{86} 040302(R) (2012). 
\bibitem{Schwartz} 
T. Schwartz, G. Bartal, S. Fishman, and M. Segev, Nature~{\bf 446}, 52 (2007).
\bibitem{SalmanOL}
S.~Karbasi, C.~R.~Mirr, P.~G.~Yarandi, R.~J.~Frazier, K.~W.~Koch, and A.~Mafi, Opt.\ Lett.\  \textbf{37}, 2304 (2012).
\bibitem{SalmanNature} 
S. Karbasi, R. J. Frazier, K. W. Koch, T. Hawkins, J. Ballato, and A. Mafi, {Nature\ Communications} \textbf{5}, 3362 (2014).
\bibitem{SegevNaturePhotonicsReview}
M.~Segev, Y.~Silberberg, and D.~N. Christodoulides, Nature Photonics \textbf{7}, 197 (2013).
\bibitem{MafiAOP}
A.~Mafi, Advances in Optics and Photonics {\bf 7} , 459--515 (2015).
\bibitem{Pendry}
J. B. Pendry, Adv. Phys. \textbf{43}, 461 (1994).
\bibitem{Berry}
M. V. Berry and S. Klein, Eur. J. Phys. \textbf{18}, 222 (1997).
\bibitem{Kondilis}
A. Kondilis and P. Tzanetakis, Phys. Rev. B \textbf{46}, 15426 (1992).
\bibitem{Bliokh}
K. Yu. Bliokh and V. D. Freilikher, Phys. Rev. B \textbf{70}, 245121 (2004).
\bibitem{Mafi}
A. Mafi, Phys. Rev. B \textbf{77}, 115140 (2008).
\end{thebibliography}
\end{document}